\documentclass[conference, 10pt]{IEEEtran}

\usepackage{cite}

\usepackage{subcaption}
\usepackage{mathtools}
\usepackage{xcolor}

\usepackage{amsmath}
\usepackage{amssymb}
\newcommand{\vt}[1]{\mathbf{#1}} 
\newcommand{\mat}[1]{\mathbf{#1}} 
 
\newcommand{\tmat}[1]{\mathbf{#1}}

\newcommand{\tr}[1]{#1\!^\mathsf{T}}

\newcommand{\p}{\vt{r}}

\newcommand{\C}{\mathbb{C}}

\newcommand{\mapinterp}{\tmat{\Lambda}}

\usepackage{url}
\usepackage{graphicx}
\usepackage{color}
\usepackage{placeins}
\usepackage{float}
\usepackage{tabularx,colortbl}
\usepackage{ifthen}
\usepackage{todonotes}

\makeatletter

\newcounter{author}
\renewcommand{\author}[2][]{
   \stepcounter{author}
   \@namedef{author@\theauthor}{#2}
   \@namedef{authorlabel@\theauthor}{#1}
}

\newcounter{address}
\newcommand{\address}[2][]{
   \stepcounter{address}
   \@namedef{address@\theaddress}{#2}
   \@namedef{addresslabel@\theaddress}{#1}
}

\newcommand{\alsep}{and}

\def\newmaketitle{\par%
  \begingroup%
  \normalfont%
  \def\thefootnote{}
  \def\footnotemark{}
  \let\@makefnmark\relax
  \footnotesize
  \footnotesep 0.7\baselineskip
  \normalsize%
  \twocolumn[\thenewmaketitle\@IEEEaftertitletext]%
  \if@IEEEusingpubid
     \enlargethispage{-\@IEEEpubidpullup}%
  \fi
  \endgroup
  \setcounter{footnote}{0}\let\maketitle\relax\let\@maketitle\relax
  \gdef\@thanks{}%
  \let\thanks\relax}

\def\thenewmaketitle{

  \newpage
  \begin{center}%
    \vskip0.2em{\Huge\@IEEEcompsoconly{\sffamily}\@IEEEcompsocconfonly{\normalfont\normalsize\vskip 2\@IEEEnormalsizeunitybaselineskip
   \bfseries\large}\@title\par}\vskip1.0em\par%
    \vspace{1ex}
    \newcounter{c@author}
    \newcounter{c@tmp}
    \ifthenelse{\value{author}=2}{%
      \newcommand{\liand}{ and }}{%
      \newcommand{\liand}{, and }}
   
    \ifthenelse{\value{address}<2}{%

      \@nameuse{author@1}%
      \stepcounter{c@author}%
      \whiledo{\value{c@author}<\value{author}}{%
        \setcounter{c@tmp}{\value{author}}%
        \addtocounter{c@tmp}{-\value{c@author}}%
        \ifthenelse{\value{c@tmp}=1}{%
          \renewcommand{\alsep}{\liand}}{\renewcommand{\alsep}{, }}%
        \stepcounter{c@author}\alsep \@nameuse{author@\thec@author}}\\%
    }
    {
      \@nameuse{author@1}${}^{(\ref{\@nameuse{authorlabel@1}})}$%
      \stepcounter{c@author}%
      \whiledo{\value{c@author}<\value{author}}{%
      \setcounter{c@tmp}{\value{author}}%
      \addtocounter{c@tmp}{-\value{c@author}}%
      \ifthenelse{\value{c@tmp}=1}{%
        \renewcommand{\alsep}{\liand}}{\renewcommand{\alsep}{, }}%
      \stepcounter{c@author}\alsep \@nameuse{author@\thec@author}%
        ${}^{(\ref{\@nameuse{authorlabel@\thec@author}})}$%
      }
    }
    \vspace{0.2ex}

    \ifthenelse{\value{address}>0}{%
      \ifthenelse{\value{address}=1}{
        {\@nameuse{address@1}}
      }
      {
        \newcounter{c@address}

        \begin{center}
        \whiledo{\value{c@address}<\value{address}}
        {
          \refstepcounter{c@address}
            ${}^{(\thec@address)}$\,%
              \label{\@nameuse{addresslabel@\thec@address}}%
              \@nameuse{address@\thec@address}\\ %
        }
        \end{center}
      } 
    }
    {
      \relax
    }
  \end{center}
}

\makeatother

\title{A Numerical Approach to Operator Filtering within the Adaptive Integral Method for Electromagnetic Integral Equations}

\author[org2]{Tommaso Pignatelli${}^{(1),}$}
\author[org1]{Viviana Giunzioni}
\author[org1]{Paolo Ricci}
\author[org4]{Matteo E. Masciocchi}
\author[org3]{\\Adrien Merlini}
\author[org1]{Francesco P. Andriulli}

\address[org1]{Department of Electronics and Telecommunications, Politecnico di Torino, 10129 Turin, Italy}
\address[org2]{Early Research Honors School, Politecnico di Torino, 10129 Turin, Italy}
\address[org3]{Microwave Department, IMT Atlantique, Brest, France}
\address[org4]{Utopia srl, 10148 Turin, Italy}

\begin{document}

\newmaketitle

\begin{abstract}
Operator filtering allows for the regularization and compression of dense integral operators, effectively mitigating the memory and computational costs associated with iterative solvers. Previous works introduced filters that leverage the analytical spectral truncation of kernels for operators of the 2D Electric Field Integral Equation (EFIE). In this contribution, we will demonstrate how to obtain filtered kernels in a discrete numerical form within the framework of an Adaptive Integral Method (AIM), yielding results entirely comparable to analytical filters. By operating directly on the discrete operator representations, the proposed strategy ensures a native and robust compatibility with fast solver schemes that analytical formulations often lack.  The effectiveness of the proposed approach will be demonstrated through numerical results, including its application to the Calderón preconditioned EFIE.
\end{abstract}

\section{Introduction}
The Electric Field Integral Equation (EFIE) is widely employed to model scattering and radiation phenomena involving perfectly electrically conducting (PEC) objects. The discretization of the EFIE using the Boundary Element Method (BEM) converts the continuous problem into a discrete matrix system~\cite{b1}. When solving electromagnetic integral equations through this approach, the high number of matrix–vector products involving large and dense matrices leads to significant memory and computational costs. Operator filtering~\cite{b2} has emerged as an impactful strategy to mitigate these requirements while improving the problem’s conditioning, as it enables the simultaneous regularization and compression of the integral operators before the solution process. Among different filtering strategies, Green’s function analytical filters have been explored, based on the truncation of a spectral representation of the operator kernel~\cite{b3}.

In this work, we propose a new discrete approach to obtain these filters without an explicit analytical formulation, relying solely on a numerical strategy. Specifically, this is done within the framework of an Adaptive Integral Method (AIM). The fundamental advantage of this numerical strategy is that it operates directly on the discrete operator representations, ensuring an intrinsic and direct compatibility with fast solver implementations that analytical formulations often lack. This contribution focuses on the operator involved in the Transverse Electric (TE) Calderón preconditioned EFIE applied to a 2D scatterer and combines theoretical considerations with numerical evidence that corroborates the theory and demonstrates the relevance of the new schemes in practice.

\section{Background and Notation}

Let $\gamma \subset \mathbb{R}^2$ be a smooth curve representing a 2D scatterer in a medium with wavenumber $k$ and impedance $\eta$. An incident field $(\boldsymbol{E}^i, \boldsymbol{H}^i)$ impinges on the object; denoting by subscript $t$ the tangential components and by $n$ the outward unit normal to the curve $\gamma$, we focus on the (TE) case. The relationship between the tangential incident field $E_t$  and the induced current $j_t$ is governed by the hypersingular operator $\mathcal{N}$ 
\begin{gather}
\eta \frac{1}{\mathrm{i} k} \mathcal{N} j_t = E_t \, ,
\end{gather}
with 
\begin{equation}
    (\mathcal{N} j_t)(\mathbf{r}) \coloneq -\frac{\partial}{\partial n} \int_\gamma \frac{\partial}{\partial n'} g(\mathbf{r}, \mathbf{r}') j_t(\mathbf{r}') \, d\mathbf{r}', 
\label{eq:N}
\end{equation}
where the Green's function is given by $g(\vt{r}, \vt{r}') \coloneq -\mathrm{j}/4 H_0^{(2)} (k|\vt{r}-\vt{r}'|)$ for $k > 0$, and \( H_0^{(2)} \) is the Hankel function of the second kind of order zero~\cite{b4}.

We discretize $j_t$ using piecewise linear Lagrange basis functions $\{\varphi_i\}_{i=1}^N$ on a mesh of $\gamma$ comprising $N$ segments of uniform length $h$, such that $j_t \approx \sum_{i=1}^N [\vt{j}_t]_i\varphi_i$. By applying the Galerkin testing scheme, we obtain the discrete system $\mathbf{N}\vt{j}_t = \vt{E}_t$.
Here, the matrix elements are $[\mathbf{N}]_{ij} = \langle \varphi_i, \mathcal{N}\varphi_j \rangle$, where $\langle f,g\rangle\coloneq \int_{\gamma}fgd\gamma$, while the excitation vector is defined as $\vt{E}_t = \langle \varphi_i, \frac{\mathrm{i}k}{\eta} E_t \rangle$.
To improve the spectral properties of the system, we employ a discrete Calderón preconditioning. Following the discretization, the preconditioned TE-EFIE is expressed as~\cite{b3}:
\begin{equation}
\mathbf{G}^{-1}\mathbf{S}\mathbf{G}^{-1}\mathbf{N}\mathbf{j}_t = \mathbf{G}^{-1}\mathbf{S}\mathbf{G}^{-1}\mathbf{E}_t,
\end{equation}
where $[\mathbf{G}]_{ij} \coloneq \langle \varphi_i, \varphi_j \rangle$ is the Gram matrix associated with the basis functions $\{\varphi_i\}_{i=1}^N$, and $\mathbf{S}$, with elements $[\mathbf{S}]_{ij} = \langle \varphi_i, \mathcal{S}\varphi_j \rangle,$ represents the discrete counterparts of the continuous single-layer operator $\mathcal{S}$  
\begin{equation}
(\mathcal{S} f)(\mathbf{r}) \coloneq \int_\gamma g(\mathbf{r}, \mathbf{r}') f(\mathbf{r}') \, d\mathbf{r}'. 
\label{eq:S}
\end{equation}

As presented in~\cite{b3}, a class of filtered operators can be derived by truncating the spectral representation of the operator's kernel \( g(\mathbf{r}, \mathbf{r}') \). 
Specifically, we define the filtered operator
    \begin{equation}
        (\mathcal{S}^\alpha j_z)(\mathbf{r}) \coloneq \int_\gamma g^\alpha(\mathbf{r}, \mathbf{r}') j_z(\mathbf{r}') \, d\mathbf{r}', \label{eq:filtered_S} 
    \end{equation}
with the modified (filtered) kernel
\begin{equation}
    g^\alpha(\mathbf{r}, \mathbf{r}') \coloneq -\frac{\mathrm{j}}{4} H_0^{(2)}(k|\mathbf{r} - \mathbf{r}'|) - \frac{1}{2\pi} \int_{s=\alpha}^{+\infty} \frac{J_0(s|\mathbf{r} - \mathbf{r}'|)s}{s^2 - k^2} \, ds, \label{eq:dynamic_kernel}
\end{equation}
where \( \alpha > k \), and \( J_0 \) is the Bessel function of the first kind of order zero~\cite{b4}.

\section{Operator Filtering via AIM: the Discrete Case}\label{sec:implementation}

In this section, we propose an effective strategy to obtain a numerically filtered discretization of the single layer operator within the framework of the AIM~\cite{b5}. This method appears particularly promising because it naturally provides access to a numerical spectrum of the Green's function (the AIM ``diagonal''), although a few caveats will have to be considered in practice.

\subsection{General Formulation}\label{sec:gen_form}

Given the set of quadrature nodes $\{\p_u\}_{u=1}^p$ and $\{\p_v\}_{v=1}^p$ with the associated weights $w_u$ and $w_v$, we compute the entry 
\begin{equation}
\begin{aligned}
 [\mat{S}]_{ij} &= \sum_{u=1}^p \sum_{v=1}^p w_u\,\varphi_i(\p_u)\,g(\p_u, \p_v)\,w_v\,\varphi_j(\p_v) \\
    &= \tr{\vt{y}} \mat{G_\mathbf{nodes}} \vt{x},
\end{aligned}\label{eq:7}
\end{equation}
where $p$ is the number of nodes required for the Gaussian quadrature, $[\vt{y}]_k = w_k\varphi_i(\p_k)$, $[\vt{x}]_k =w_k\varphi_j(\p_k)$, and $\mat{G_\mathbf{nodes}}$ is the dense interaction matrix with entries $[\mat{G_\mathbf{nodes}}]_{i,j} = g(\p_i, \p_j)$. The above expression~\eqref{eq:7} is to be understood in a formal sense, as some matrix entries involve singular integrals and require special numerical treatment. While this issue is typically addressed by employing proper singularity extraction or cancellation strategies, in the specific case of filtering, a simpler approach can be adopted, as delineated below.

We define a regular Cartesian grid of dimension $2L \times 2L$ centered at the origin, encompassing the scatterer $\gamma$. The grid consists of $M = m^2$ total points, with $m\in \mathbb{N}$, and a uniform grid step $\varepsilon$ along both axes, such that $\varepsilon = 2L / (m-1)$, where $m$ represents the number of points per side. Then, we approximate the matrix $\mathbf{G_{nodes}}$ by its factorization $\mat{G_\mathbf{nodes}} \approx \tr{\mapinterp} \mathbf{G_{grid}} \mapinterp$, where $\mapinterp \in \mathbb{C}^{M\times Np}$ is a sparse interpolation matrix that projects the quadrature nodes onto the grid points, and $\mathbf{G_{grid}} \in \mathbb{C}^{M\times M}$ is a Toeplitz matrix representing the kernel interactions on the uniform grid points.

To efficiently compute the product $\mathbf{G_{grid}} \vt{u}$ (where $\vt{u} = \mapinterp \vt{x}$), as is standardly done in AIM, we embed the Toeplitz matrix into a circulant matrix to utilize the Fast Fourier Transform (FFT) algorithm. We define $\vt{g_{grid}} \in \C^{m\times m}$ the first row of $\mathbf{G_{grid}}$, reshaped as a square matrix, and $\vt{c_{grid}} \in \C^{(2m-1)\times(2m-1)}$ the matrix  created by mirroring $\mathbf{g_{grid}}$ about its last column and row. Finally, $\vt{b} \in \C^{(2m-1)\times(2m-1)}$ is constructed by taking $\vt{u}$, reshaping it as a square matrix in $\C^{m\times m}$, and padding it with zeros to match the dimensions of $\mathbf{c_{grid}}$.

Finally, the matrix-vector product is computed with  $\mathcal{O}(M^{1.5})$ complexity~\cite{b6} using element-wise multiplication ($\cdot{*}$) in the frequency domain:
\begin{equation}
    \mat{G_\mathbf{nodes}}\vt{x} \approx \tr{\mapinterp} \left[ \mathcal{F}^{-1} \left( \mathcal{F}(\vt{c_{grid}}) \cdot{*} \mathcal{F}(\vt{b}) \right) \right]_{\text{restr}}
    \label{eq:Toepl}
\end{equation}
where $[\cdot]_{\text{restr}}$ indicates removing the padding to return to the original grid size $M$, and $\mathcal{F}$ is the 2D discrete Fourier Transform.

\subsection{Filtering}

\begin{figure}
    \centering
    \begin{subfigure}[b]{2.4in} 
        \centering
        \includegraphics[width=\columnwidth]{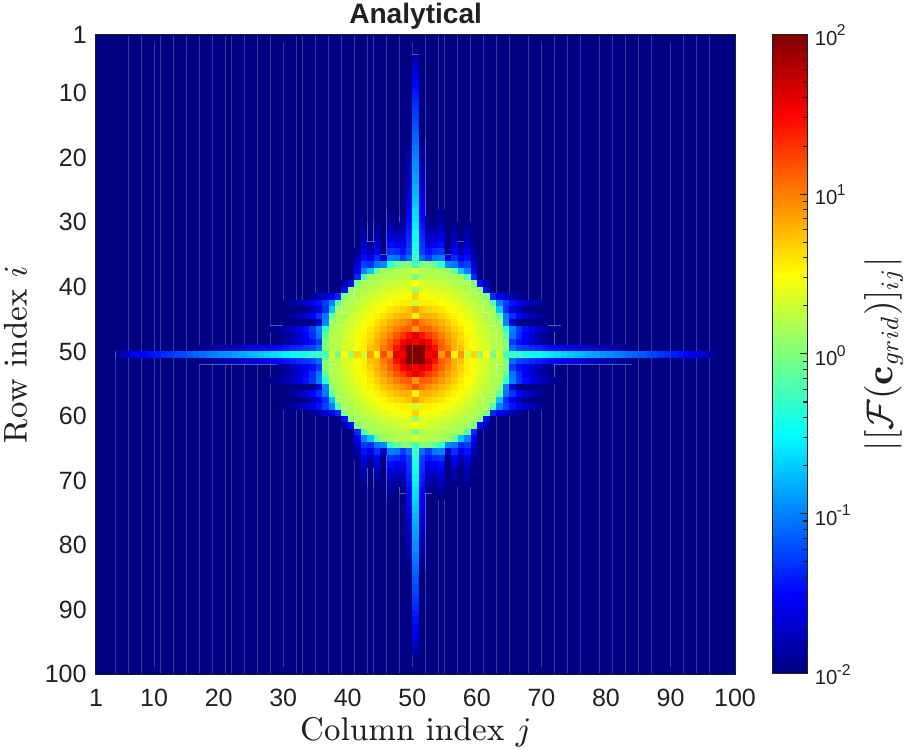} 
        \caption{}
        \label{fig:1a}
    \end{subfigure}

    \begin{subfigure}[b]{2.4in}
        \centering
        \includegraphics[width=\columnwidth]{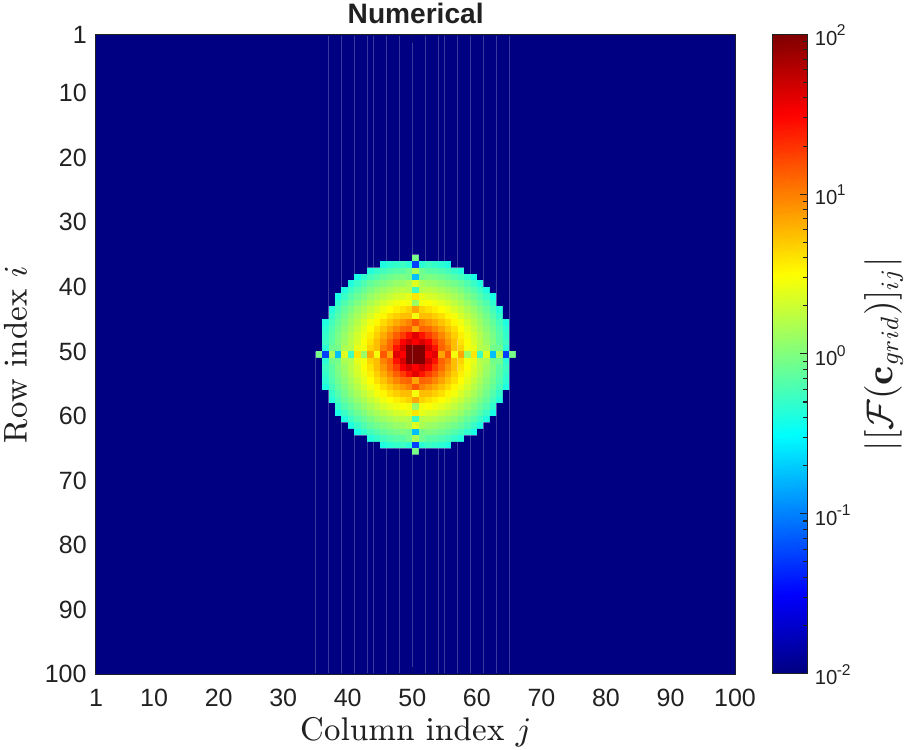}
        \caption{}
        \label{fig:1b}
    \end{subfigure}

    \caption{Comparison between $\mathcal{F}(\vt{c_{grid}})$ where (a) an analytically filtered kernel and (b) our numerical approach have been used. The mesh used is a circle, $m=50$.}
    \label{fig1}
\end{figure}

We observe that, since $\vt{g_{grid}}$ contains the interactions between the upper-left vertex of the Cartesian grid and the $M$ points of the grid, $\vt{c_{grid}}$ appropriately reshaped can be seen as the Green's function evaluated at discrete points on the 2D plane. 
Since the analytical filter has the effect of a circular truncation with radius $\alpha$ in the Fourier Transform domain, we would like to reproduce it numerically in $\mathcal{F}(\vt{c_{grid}})$, but first we have to notice a potential issue.
By Fourier transforming the analytically filtered Green's function kernel~\eqref{eq:dynamic_kernel} over a limited domain (the grid), we obtain:
\begin{equation}
    \mathcal{F}\left(g^{\alpha}\Pi_{L}(x,y)\right)(k_x,k_y) = \mathcal{F}(g^{\alpha}) * \text{sinc}(k_x L, k_y L),
\end{equation}
where $\Pi_{L}$ is the square window of side $2L$ centered at the origin and $\text{sinc}(u,v) = \text{sinc}(u) \cdot \text{sinc}(v) = \frac{\sin(u)}{u}\frac{\sin(v)}{v}$. However, since the impact of this convolution on the overall frequency spectrum can be made negligible with a sufficiently large grid, we can directly enforce a numerical filtering by circularly truncating $\mathcal{F}(\vt{c_{grid}})$ (which is already computed and available from the AIM implementation \eqref{eq:Toepl}) at the same  cutoff used in the analytical filters, obtaining a filtered version $\mathcal{F}(\vt{c_{grid}}^\alpha)$ (Fig.~\ref{fig1}).

Finally,  the discretized filtered operator $\mat{S}^\alpha$ is  obtained as
\begin{equation}\label{eq:S num}
    [\mat{S}^\alpha]_{i,j} = \tr{\vt{y}} \tr{\mapinterp} \left[ \mathcal{F}^{-1} \left( \mathcal{F}(\vt{c_{grid}}^\alpha) \cdot{*} \mathcal{F}(\vt{b}) \right) \right]_{\text{restr}}.
\end{equation}

\subsection{Singularity}
We now address the kernel singularity at the origin. Since the target  filtered Green’s function has a filtered spectrum at infinity (i.e., for high spatial frequencies), the resulting spatial expression is regular at the origin. 
For this reason, in the numerical implementation, the singularity at the origin is handled by just setting the self-interaction term in $\mathbf{G_{grid}}$ to zero. We demonstrate that this does perturb the final result in a controllable way. Since the Green's function depends solely on the vector difference $\mathbf{r} - \mathbf{r}'$, we adopt the notation $g(x,y)$ (thus setting $\mathbf{r} - \mathbf{r}'=(x,y)$) to represent the kernel as a function of the relative coordinates. In fact, if we set to zero the value of the 2D Green's function in the central square of side $\varepsilon$, as in
$
g_\varepsilon(x,y)=g(x,y)-g(x,y)\,\Pi_\varepsilon(x,y)\,,
$
by linearity and the convolution theorem
\begin{equation}
\mathcal{F}\{g_\varepsilon\}
=\mathcal{F}\{g\}
-\mathcal{F}\{g\}\ast\mathcal{F}\{\Pi_\varepsilon(x,y)\}.
\end{equation}
Since 
$\mathcal{F}\{\Pi_\varepsilon(x,y)\}(k_x,k_y)
=\varepsilon^2\,\mathrm{sinc}(\tfrac{k_x\varepsilon}{2})\,\mathrm{sinc}(\tfrac{k_y\varepsilon}{2})
= \varepsilon^2+o(\varepsilon^2)$ for \(\varepsilon\to0\),
one obtains
\begin{equation}
\mathcal{F}\{g_\varepsilon\}
=\mathcal{F}\{g\}\bigl(1-\varepsilon^2\bigr)+o(\varepsilon^2)
\;\xrightarrow[\varepsilon\to0]{}\;
\mathcal{F}\{g\}.
\end{equation}
This confirms that the Fourier transform remains unchanged in the \(\varepsilon\to0\) limit, which corresponds to the case where $M\rightarrow\infty$.

\section{Numerical Results}\label{sec:num_res}

\begin{figure}
\centerline{\includegraphics[width=\columnwidth]{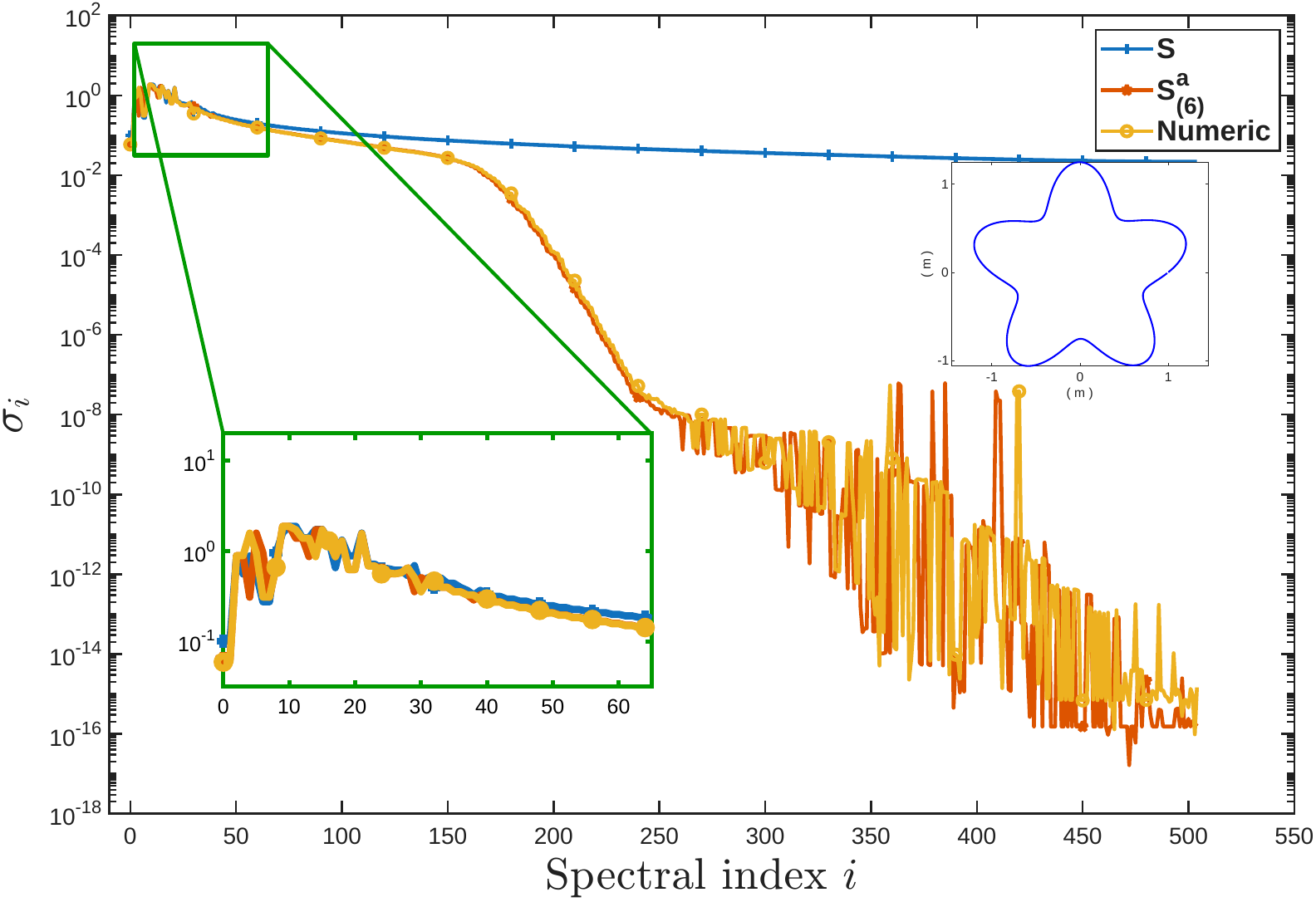}}
\caption{Comparison of the singular values of $\mat{S}$, $\mat{S}^\alpha$ using \eqref{eq:dynamic_kernel} and S filtered through the AIM numerical approach, ordered by the singular vectors of the Laplace-Beltrami operator on the mesh illustrated in the figure.}
\label{fig2}
\end{figure}

We now proceed to validate the effectiveness of the proposed numerical filter. Figure~\ref{fig2} illustrates the impact of the different filtering approaches on the spectrum of the discretized single layer operator $\mat{S}$. The numerical analysis was performed at $f=0.4$ GHz, with a mesh size $h=\pi/{15k}$, $p=10$, using an AIM grid with $m=600$, $L = 1.2$ m, a bilinear interpolation order of 10 for $\mapinterp$, and $\alpha = 8k$.
It can be observed that both the analytical \eqref{eq:dynamic_kernel} and the numerical filter \eqref{eq:S num} produce a comparable effect on the operator's spectrum, which, as expected, exhibits an exponential decay after the filtering point.

\begin{figure}
\centerline{\includegraphics[width=\columnwidth]{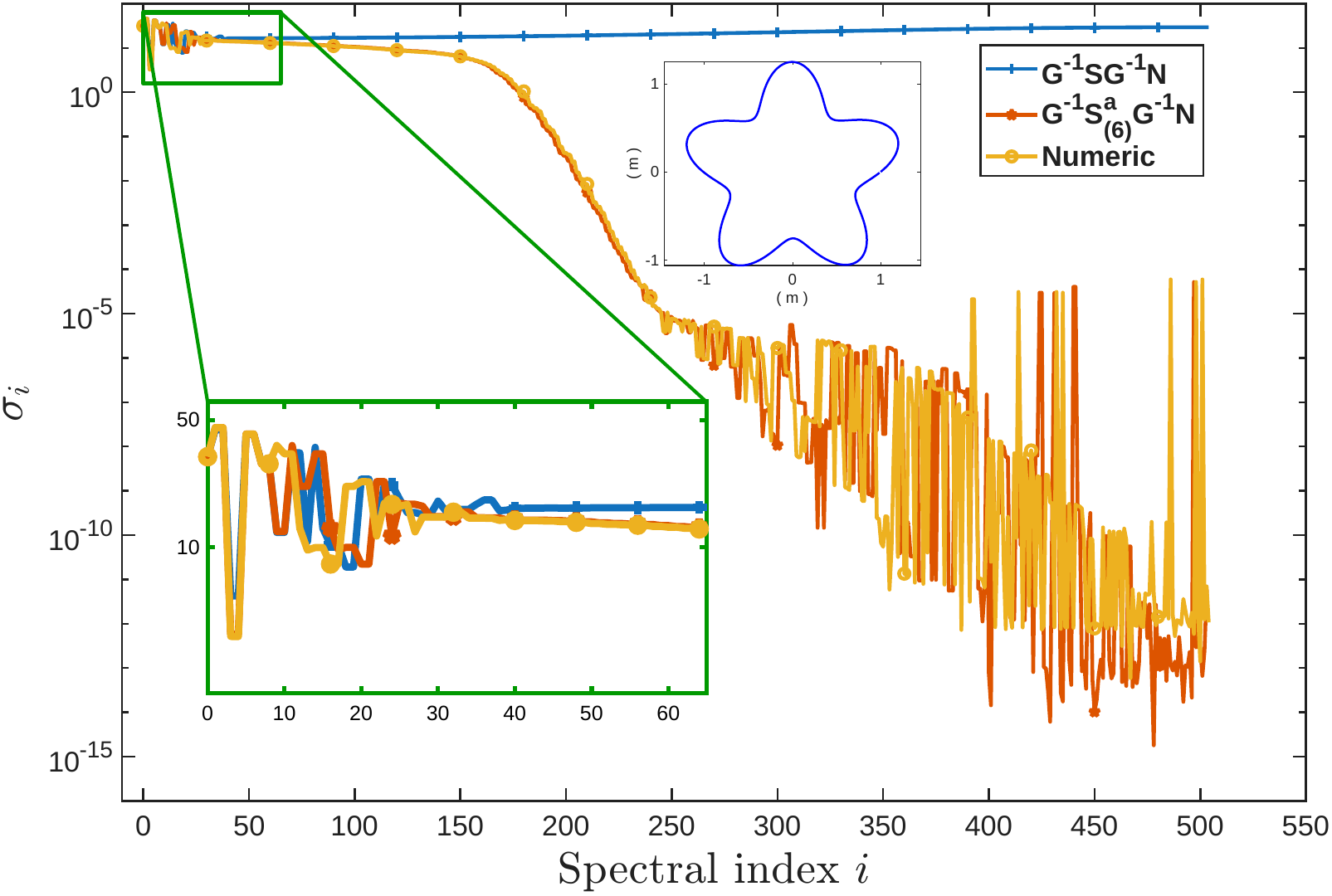}}
\caption{Comparison of the singular values of $\mat{G^{-1}SG^{-1}N}$ and $\mat{G^{-1}S^{\alpha}G^{-1}N}$ where $\mat{S^{\alpha}}$ is computed using  \eqref{eq:dynamic_kernel} and the AIM numerical approach, ordered by the singular vectors of the Laplace-Beltrami operator on the mesh illustrated in the figure.}
\label{fig3}
\end{figure}

With these filtered operators, we can finally evaluate the effectiveness of the filtering procedure in the Calderón preconditioned TE-EFIE.
Figure~\ref{fig3}, in particular, shows the singular values of $\mat{G}^{-1} \mat{S} \mat{G}^{-1} \mat{N}$ compared to the singular values of $\mat{G}^{-1} \mat{S}^{\alpha} \mat{G}^{-1} \mat{N}$, where $\mat{S}^{\alpha}$ is computed using both the analytical \eqref{eq:dynamic_kernel} and the AIM numerical approach \eqref{eq:S num}.
As the objective is limited to preconditioning, our focus remains on the general spectral distribution and its eventual decay, rather than on fine-scale details. Accordingly, numerical results demonstrate the effectiveness and applicability of the approach in this case.

\section*{ACKNOWLEDGEMENT}
The work of this paper has received funding from the European Innovation Council (EIC) through the European Union’s Horizon Europe research Programme under Grant 101046748 (Project CEREBRO).

\end{document}